\documentclass{aastex62}

\submitjournal{AJ}

\shorttitle{NGC\,1978}
\shortauthors{A.E. Piatti \& J. Bailin}

\begin{document}

\title{Search for  a metallicity spread in the multiple population Large Magellanic Cloud cluster NGC\,1978}

\author[0000-0002-8679-0589]{Andr\'es E. Piatti}
\affiliation{Consejo Nacional de Investigaciones Cient\'{\i}ficas y T\'ecnicas,  Godoy Cruz 2290, C1425FQB,
Buenos Aires, Argentina}
\affiliation{Observatorio Astron\'omico de C\'ordoba, Laprida 854, 5000, C\'ordoba, Argentina}
\correspondingauthor{Andr\'es E. Piatti}
\email{e-mail: andres@oac.unc.edu.ar}

\author{Jeremy Bailin}
\affiliation{Department of Physics and Astronomy, University of Alabama, Box 870324, Tuscaloosa, AL, 
35487-0324 USA}
\affiliation{Steward Observatory, University of Arizona, 933 N Cherry Ave., Tucson, AZ, 85721-0065 USA}

\begin{abstract}
We report on the spread of [Fe/H] values in the massive Large Magellanic Cloud cluster
NGC\,1978, recently confirmed to harbor multiple populations of nearly the same age.
We used accurate Str\"omgren photometry of carefully selected cluster red giant branch stars
along with a high-dispersion spectroscopy-based calibration of the metallicity-sensitive index $m_1$.
Once accounting for the photometry quality, assessed from extensive artificial stars tests
to trace the photometric uncertainties as a function of the position to the cluster's center
as well as the stellar brightness, and those from the metallicity calibration, we found that
NGC\,1978 exhibits a small metallicity spread of 0.035 dex ($\pm$ 0.019-0.023),
depending on whether stars with individual $\sigma$[Fe/H] $\le$ 0.15 dex or those located in the
cluster's outer areas are considered. Such a spread in [Fe/H] is consistent with a
cluster formation model with self-enrichment, if mass loss higher than 90 per cent
due to stellar evolutionary and galactic tidal effects is assumed. Nevertheless, scenarios where the
apparent [Fe/H] variation reflects CN abundance anomalies or less extreme mass loss models 
with environmentally-dependent self enrichment should not be ruled out.
\end{abstract} 


\section{Introduction}

A number of theoretical models have recently proposed different scenarios
to explain anomalies in the abundances of several chemical elements in massive
clusters harboring multiple populations (MPs)
\citep[see, e.g.][and references therein]{bt2016,bekki2017,bailin2018,gielesetal2018,kimeal2018}.
They have been mainly stimulated by observational results of
anti-correlations between chemical abundances of light elements.
Some of the models also suggest mechanisms to obtain
intrinsic [Fe/H] spreads $>$ 0.05 dex, as observed at the present time in 8  Milky Way (MW) old 
globular clusters \citep{marinoetal2015,limetal2017}.

NGC\,1978 is a massive \citep[$\sim$3$\times$10$^5$M$_\odot$][]{pb16b}
intermediate-age Large Magellanic Cloud (LMC) cluster \citep[2 Gyr][]{mietal09,goudfrooijetal14}
with a mean metallicity of [Fe/H] = -0.38$\pm$0.05 dex \citep{ferraroetal2006}.
\citet{ledereretal2009} and \citet{davidge2018} have discussed its
puzzling metallicity pattern overviewing previous studies carried out on this object,
although they did not discuss any possible Fe-abundance spread.
NGC\,1978 has recently been found to harbor MPs based on [N/Fe] variations among red giant branch 
(RGB) stars \citep{martocchiaetal2018a}, a feature also seen in almost all MW globular 
clusters with MPs. Hence, the authors concluded that there must be a common formation 
and evolutionary process for massive clusters ($\ga$ 10$^5$M$_\odot$), regardless of their
ages. Additionally, NGC\,1978's MPs are of nearly the same age 
\citep{martocchiaetal2018b}, so their abundance dispersions have somehow originated concurrently.

In this paper we report on the magnitude of the intrinsic [Fe/H] spread in 
NGC\,1978. In order to measure it, we made use of the Str\"omgren photometric data set described
in Section 2. We performed a careful selection of cluster RGB members and estimated
individual [Fe/H] values from a high-dispersion spectroscopy-based calibration of the
metallicity sensitive $m_{\rm 1}$ index, as detailed in Section 3. In Section 4 we 
analyze the metallicity estimates in the light of a maximum likelihood approach and discuss
the results in the context of self-enrichment cluster formation models. Finally, Section
5 summarizes the main conclusions of this work.

\section{Str\"omgren photometry data set}

We downloaded publicly available Str\"omgren $vby$ images from the National Optical Astronomy 
Observatory (NOAO) Science Data Management (SDM) 
Archives\footnote{http //www.noao.edu/sdm/archives.php.}. They were obtained, along with
suitable calibration and standard field images, as part of an observational program aiming at
studying the chemical evolution of the LMC from clusters and field stars (program ID: SO2008B-0917, PI: Pietrzynski). 
The images were obtained on the night of 18 January 2008
with the SOAR Optical Imager (SOI) mounted on the 4.1m Southern Astrophysical Research 
(SOAR) telescope (FOV= 5.25$\arcmin$$\times$5.25$\arcmin$, scale=0.154$\arcsec$/px), and are
of excellent quality (typical FWHM $\sim$ 0.6$\arcsec$). They consist of one exposure of 
350, 140, and 90 sec in the $v$, $b$, and $y$ filters, respectively. In order to properly process them,
we followed the SOI's pipeline guidance described at 
http://www.ctio.noao.edu/soar/content/soar-optical-imager-soi.

To standardize our photometry we measured instrumental $vby$ magnitudes of the standard
stars HD\,64, HD\,3417, HD\,12756, HD\,22610, TYC\,7583-1622-1, TYC\,8104-969-1, HD\,57568 and
HD\,58489 \citep{hm1998,p2005}. The stars were observed at airmass between 1.05 and 2.83
along the whole night.  Each star was observed twice at a given airmass, in order to
place it in each of the two CCDs used by SOI. We then performed fits of the expressions:\\

$v = v_1 + V_{\rm std} + v_2\times X_v + v_3\times (b-y)_{\rm std} + v_4\times m_{\rm 1 std}$,\\

$b = b_1 + V_{\rm std} + b_2\times X_b + b_3\times (b-y)_{\rm std}$,\\

$y = y_1 + V_{\rm std}  + y_2\times X_y + y_3\times (b-y)_{\rm std}$,\\

\noindent where  $v_i$, $b_i$ and $y_i$ are the i-th fitted 
coefficients, and $X$ represents the effective airmass. Table~\ref{tab:table1} shows
the resulting coefficients for stars measured in the two different CCDs, separately. 
As can be seen, there is an excellent agreement between the independent transformation
coefficients from both CCDs. For this reason, we decided to use all the measured
stars, regardless their positions in SOI. The resulting coefficient are also
listed in Table~\ref{tab:table1}.

\begin{deluxetable*}{ccccccc}
\tablecaption{Str\"omgren transformation coefficients.\label{tab:table1}}
\tablehead{\colhead{Filter} & \colhead{coef$_{\rm 1}$} & \colhead{coef$_{\rm 2}$} & \colhead{coef$_{\rm 3}$} & \colhead{coef$_{\rm 4}$} & \colhead{rms}}
\startdata
$v$    & $\#$ 1 & 1.255$\pm$0.023& 0.276$\pm$0.012& 1.823$\pm$0.087& 1.148$\pm$0.095 &  0.009 \\
       & $\#$ 2 & 1.264$\pm$0.027& 0.290$\pm$0.015& 1.822$\pm$0.103& 1.135$\pm$0.111 &  0.011 \\
       & full   & 1.259$\pm$0.023& 0.280$\pm$0.012& 1.823$\pm$0.087& 1.141$\pm$0.095 &  0.009 \\
       &        &         &      &        &       &        \\
$b$    &  $\#$ 1& 0.998$\pm$0.016& 0.181$\pm$0.008 &0.940$\pm$0.027  &       &0.009        \\ 
       &  $\#$ 2& 1.009$\pm$0.010& 0.188$\pm$0.005 &0.928$\pm$0.016  &       &0.006        \\
       &  full  & 1.003$\pm$0.010& 0.184$\pm$0.005 &0.934$\pm$0.016  &       &0.006        \\
       &        &         &         &        &       &        \\
$y$    &  $\#$ 1&0.982$\pm$0.024    &0.131$\pm$0.013 &-0.024$\pm$0.041  &       &0.014        \\ 
       &  $\#$ 2&0.999$\pm$0.015    &0.131$\pm$0.008 &-0.007$\pm$0.025  &       &0.009        \\
       &  full  &0.990$\pm$0.015    &0.131$\pm$0.008 &-0.015$\pm$0.025  &       &0.009        \\\hline
\enddata
\end{deluxetable*}

\begin{figure}
     \includegraphics[width=\columnwidth]{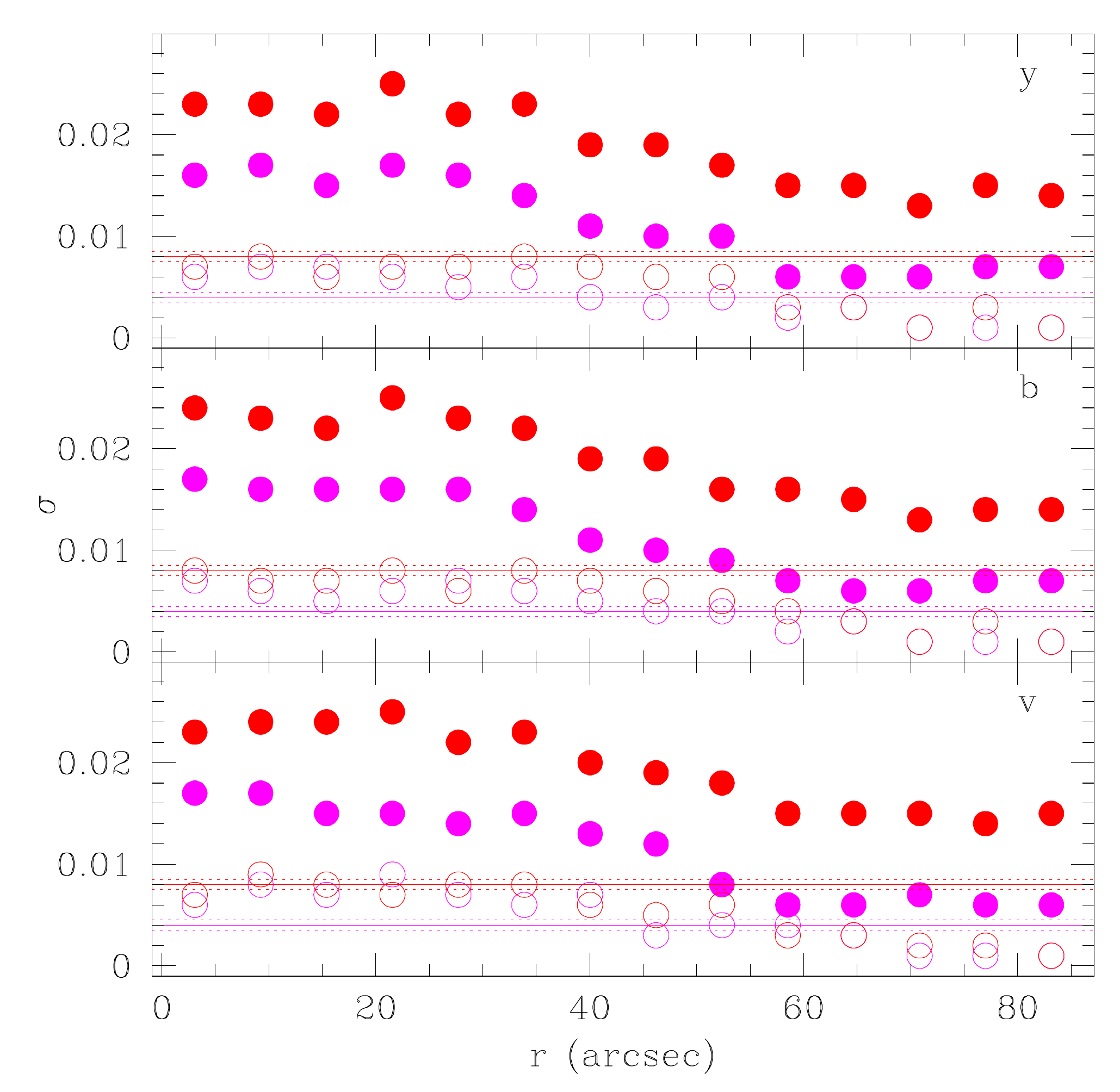}
   \caption{Photometric error estimates for the  $vby$ filters as a function of the distance
to the cluster center. Magenta and red symbols are for $V$ = 16.5 and 18.5 mag, respectively. Solid
and dotted lines represent the errors returned from applying the PSF models. Open and filled
circles are the difference between the mean input and output magnitudes of the thousand realizations of
artificial star tests and their rms errors, respectively.}
 \label{fig1}
\end{figure}

We obtained instrumental $vby$ magnitudes of stars located in the field of NGC\,1978 using
the stand-alone versions of  {\sc daophot}, {\sc allstar}, {\sc daomatch} and
{\sc daomaster} routine packages \citep{setal90}. The magnitudes were derived from
point-spread-function (PSF) fits performed using previously generated spatially
quadratically varying PSFs. These PSFs were modelled  from a sample of nearly 100 interactively 
selected stars distributed throughout the image, previously cleaned from fainter 
contaminating neighbouring 
stars using a preliminary PSF built with nearly 40 relatively bright, well-isolated stars. 
Once we applied the resulting PSF to an image, we took advantage of the subtracted image to
identify new fainter sources which were added to the final photometric catalogue. In each of the
three iterations performed, we did the PSF photometry for the whole sample of identified sources.
Finally, we transformed the instrumental magnitudes into the Str\"omgren photometric system
using the coefficients listed in Table~\ref{tab:table1}.  Fig.~\ref{fig1} illustrates the mean
and standard dispersion (solid and dotted lines) of the resulting photometric errors 
for two different magnitude levels, namely $V$ = 16.5 mag and 18.5 mag, respectively. 
These magnitudes
roughly correspond to the upper and lower limits of the cluster RGBs used in this work
(see Fig.~\ref{fig2}). Note that these errors do not increase as function of radius,
revealing that the PSF model is accurate. Table~\ref{tab:table2}  shows a portion of the
resulting photometric catalog.

We rigorously examined the quality of our photometry aiming at obtaining robust estimates of the photometric 
 errors. In doing that, we performed the widely accepted artificial star tests 
 \citep[see, e.g.][and references therein]{pb16a,pc2017,pm2018} by using the stand-alone {\sc addstar} program 
in the {\sc daophot} package \citep{setal90} to add synthetic stars, 
generated bearing in mind the color and magnitude distributions 
of the stars in the color-magnitude diagram (CMD) as well as the cluster radial stellar 
density profile. We added a number
of stars equivalent to $\sim$ 5$\%$ of the measured stars in order to avoid in the synthetic images 
significantly 
more crowding than in the original images. On the other hand, to avoid small number statistics in the
 artificial-star 
analysis, 
we created a thousand different images for each original one. We used the option of entering the number of
 photons
per ADU in order to properly add the Poisson noise to the star images. 

We then repeated the same steps to obtain the photometry of the synthetic images as described above, 
i.e., 
performing three passes with the {\sc daophot/allstar} routines. 
Fig.~\ref{fig1} shows with open circles the difference between the mean input and output magnitudes 
for the thousand realizations of added synthetic stars - matched using the {\sc daomatch} and 
{\sc daomaster} tasks - for the magnitude
levels 16.5 mag and 18.5 mag, respectively. These differences show that we were able to recover
reliable magnitudes along the whole range of distances. Nevertheless, as
a conservative representative of the photometry uncertainties we adopted the
rms errors of all artificial star tests (filled circles in Fig~\ref{fig1}), 
that best reflect the crowding effects, as expected. 

\begin{deluxetable*}{cccccccccc}
\tablecaption{Str\"omgren photometry of stars in the field of NGC\,1978.\label{tab:table2}}
\tablehead{\colhead{Star} & \colhead{R.A.} & \colhead{Dec.} & \colhead{$r^a$} &\colhead{$V$} & \colhead{$\sigma(V)$} & \colhead{$b-y$} &
\colhead{$\sigma(b-y)$} & \colhead{$m_{\rm 1}$} & \colhead{$\sigma(m_{\rm 1})$} \\
\colhead{ } & \colhead{(deg)} & \colhead{(deg)} & (arcsec) & \colhead{(mag)} & \colhead{(mag)} & \colhead{(mag)} &
\colhead{(mag)} & \colhead{(mag)} & \colhead{(mag)}}
\startdata
  --- & ---      & ---      & ---    & ---   &  ---   & ---   & ---    & ---\\
  832 & 82.180946&   -66.205498  & 110.2& 18.081 & 0.016 &  0.699 & 0.021 &  0.458 & 0.034\\ 
  833 & 82.208344&   -66.215805  & 79.0& 19.133 & 0.013 &  0.611 & 0.019 &  0.340 & 0.031\\ 
  834 & 82.148354&   -66.191719 & 169.0 & 18.132 & 0.008 &  0.618 & 0.012 &  0.350 & 0.019\\ 
  --- & ---      & ---      & ---    & ---   &  ---   & ---   & ---    & ---\\
\enddata
\tablenotetext{a}{Distance to the cluster center.}
\tablecomments{A portion of the table is presented here for guidance of its contents. The entire table is available as Supplementary 
material in the online version of the journal.}
\end{deluxetable*}

\section{metallicity estimates}

To estimate the metallicity of each star we used the calibration of giant stars based on both the 
metallicity-sensitive index $m_{\rm 1}$ and the $v-y$ color derived by
\citet{calamidaetal2007}. Particularly, we used the inverted expression, i.e., [Fe/H] as
a function of the reddening free $m_{\rm 1}$$_o$ and $(v-y)_o$ colors 
\citep[eq. (3)]{franketal2015}, as follows:

\begin{equation}
{\rm [Fe/H]} = \frac{{m_{\rm 1}}_o + a_1(v-y)_o + a_2}{a_3 (v-y)_o + a_4},
\end{equation}\\

\noindent where $a_1$ = -0.521$\pm$0.001, $a_2$=0.309, $a_3$=0.159$\pm$0.001 and
$a_4$=-0.090$\pm$0.002, respectively. 
In doing that, we first adopted the $E(B-V)$ value used by 
\citet{martocchiaetal2018a,martocchiaetal2018b} and the $E(X)/E(B-V)$ ratios given by 
\citet{cm1976} to obtain $E(m_{\rm 1})$ and $E(v-y)$ color excesses, from which we corrected
 $m_{\rm 1}$ and $v-y$ for the effects of reddening. Because the \citet{calamidaetal2007}
calibration is based on old globular clusters, and NGC\,1978 is much younger,  
the derived [Fe/H] distribution result is centerd on a more metal-poor value 
([Fe/H] = -1.0 dex) than the cluster mean metallicity ([Fe/H] = -0.38 dex) 
\citep{ferraroetal2006} due to 
the well-known 
age effect on metallicities obtained from the cluster RGB \citep[see, e.g.][]{getal03,os15}. 
Nevertheless, that  constant offset does not
have any effects on the subsequent analysis, since we are interested
only in the spread of [Fe/H] values with respect to the mean value.
Because the MPs of NGC\,1978 have essentially identical ages \citep{martocchiaetal2018b},
the difference in ages will not artificially inflate the measured [Fe/H] dispersion.

In order to estimate the uncertainties of the individual [Fe/H] values we thoroughly performed 
a full analytical propagation of errors, including those on the
calibration coefficients of eq. (1), as follows:\\

$\sigma({\rm [Fe/H]})^2 = \left(\frac{\partial {\rm [Fe/H]}}{\partial a_1} \sigma(a_1)\right)^2 
+ \left(\frac{\partial {\rm [Fe/H]}}{\partial a_2} \sigma(a_2)\right)^2 +
\left(\frac{\partial {\rm [Fe/H]}}{\partial a_3}\sigma(a_3)\right)^2 + 
  \left(\frac{\partial {\rm [Fe/H]}}{\partial a_4}\sigma(a_4)\right)^2 + 
  \left(\frac{\partial {\rm [Fe/H]}}{\partial {m_{\rm 1}}_o}\sigma({m_{\rm 1}}_o)\right)^2 + 
\left(\frac{\partial {\rm [Fe/H]}}{\partial (v-y)_o}\sigma((v-y)_o)\right)^2$,\\

\vspace{0.5cm}

$\sigma({\rm [Fe/H]})^2 =  
\left(\frac{0.001(v-y)_o}{c}\right)^2 + \left(\frac{0.001{\rm [Fe/H]}(v-y)_o}{c}\right)^2 +
\left(\frac{0.002{\rm [Fe/H]}}{c}\right)^2 +
\left(\frac{\sigma({m_{\rm 1}}_o)}{c}\right)^2 + \\
\left(\frac{(-0.521c - 0.159({m_{\rm 1}}_o +0.309 -0.521(v-y)_o)\sigma((v-y)_o)}{c^2}\right)^2$

\vspace{0.5cm}

\noindent where c = $a_3(v-y)_o + a_4$, and $\sigma({m_{\rm 1}}_o)$ and $\sigma((v-y)_o)$
are the photometric errors in ${m_{\rm 1}}_o$ and $(v-y)_o$, respectively, according to
the position of the stars with respect to the cluster's center (see Fig.~\ref{fig1}) and
the expressions: \\

 $(v-y)_o = v-y - 1.67\times 0.74E(B-V),\\
$

$\sigma((v-y)_o)^2 = \sigma(v)^2 + \sigma(y)^2 + (1.67\times 0.74\sigma(E(B-V)))^2 \\
$

\noindent and\\

$ {m_{\rm 1}}_o= (v-b) - (b-y) + 0.33\times 0.74E(B-V),\\
$

$\sigma({m_{\rm 1}}_o)^2 = \sigma(v)^2 + 4\sigma(b)^2 + \sigma(y)^2 + (0.33\times 0.74\sigma(E(B-V)))^2.\\
$

The RGB star sample we finally kept to investigate any metallicity variation in NGC\,1978 was
carefully selected following these criteria:

$\bullet$ the stars are located inside the cluster tidal radius \citep{wz11,pb16b}.

$\bullet$ stars located beyond 0.05 mag in $b-y$ from the cluster RGB ridge line 
in the $V$ versus $b-y$ CMD were
discarded. Note that the cluster RGB in the $V$ versus $b-y$ CMD is not strongly affected
by metallicity, so it results in a narrow star sequence.
We only considered RGB stars brighter than the cluster red clump/horizontal branch.

According to \citet{kamathetal2010}, NGC1978 hosts a non-zero population of AGB
stars, which can add a contamination of up to 10$\%$ of the stellar sample
as analyzed here. To assess their influence on the metallicity spread, we
first consulted a set of Padova isochrones
\citep{betal12}, available in the Str\"omgren system, with ages and
metallicities tailored to NGCº,1978.
We then created random samples of stars matching the observed numbers. and
designed
them to lie either on the theoretical RGB or AGB tracks;
next, we computed their metallicities from the above calibrations. 
Despite the inherent mono-metallicity of the isochrones, this procedure introduces a
standard deviation on the order of 0.04 dex from theoretical considerations for both
classes of stars. Moreover, metallicities derived from AGB stars are on average lower by
merely 0.02
dex, so that the inclusion of a minor contamination in our presumed RGB
sample,
would not lead to a significant inflation of the intrinsic dispersion if
they were
misclassified as RGB stars. For instance, from V = 16.6 down to 17.0 7 out
of 9 selected stars are
RGB stars observed from high-dispersion spectroscopy \citep{ferraroetal2006}, that
expand the readily visible range of metallicities in NGC\,1978. From V =
17.5 down to
18.4, the sequence of AGB star is clearly distinguished from that of RGB
stars.
Moreover, we note that, as AGB stars are warmer and less massive than RGB
stars, their
metallicities should follow different relations of the
\citet{calamidaetal2007} calibration that used here, also due to the
different
strength of the molecular bands, with weaker CN- and CH-bands. Indeed,
four AGB
stars found in
our sample have [Fe/H] values much more different that those shown by RGB
stars.

\begin{figure*}
   \includegraphics[width=\textwidth]{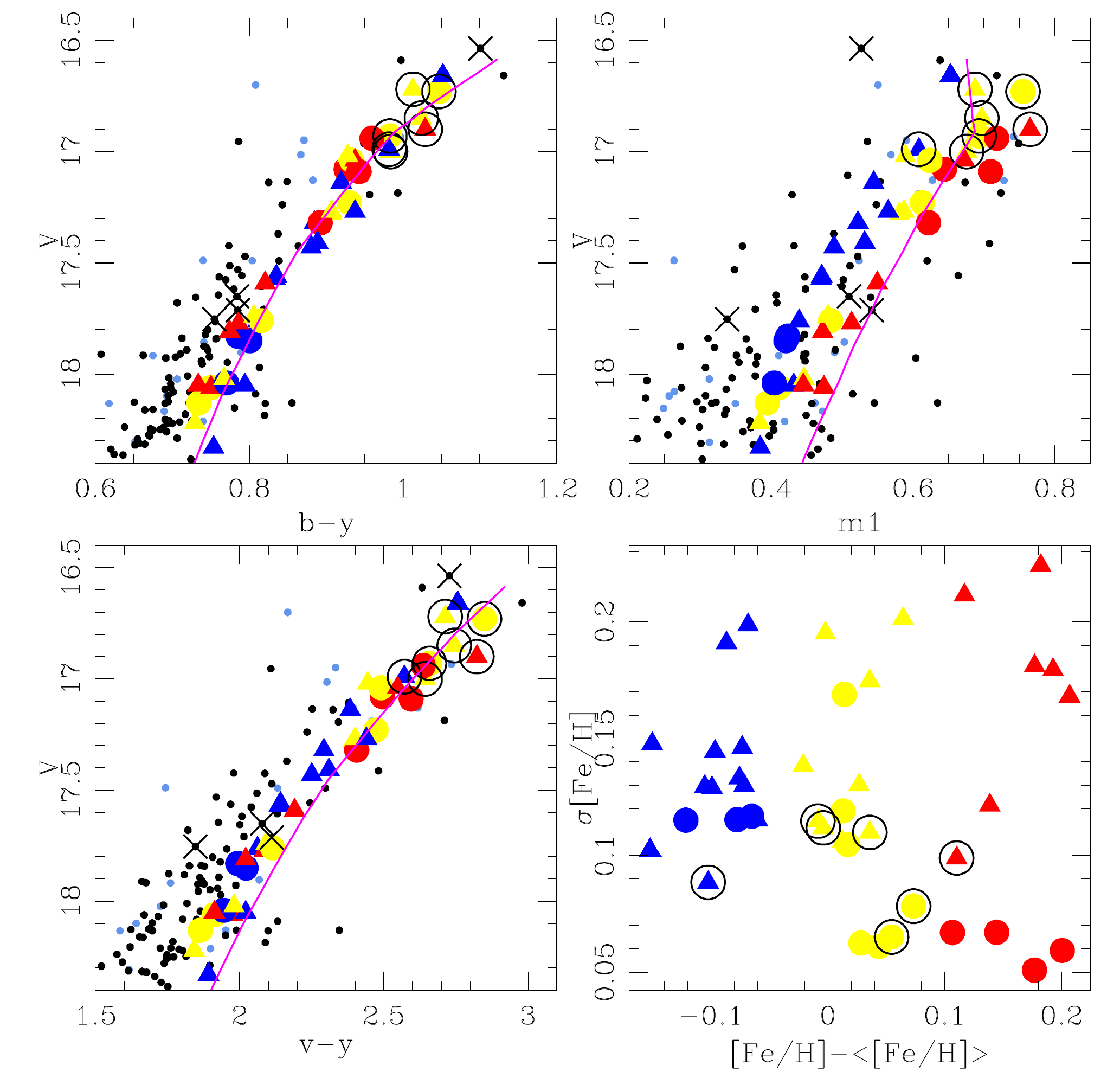}
    \caption{Three different CMDs for measured stars distributed  within the cluster region (black
small filled circles) and those selected for the metallicity analysis (large
    filled circles and triangles, color coded by metallicity as illustrated in the bottom-right panel),
    and
those for an equal cluster area star field (light-blue small filled circles).
Crosses represent AGB stars measured in the cluster field \citep{kamathetal2010}.
Large filled triangles and circles correspond to stars located closer and farther than 50 arcsec from the
cluster's center. 
The bottom-right panel shows the [Fe/H] difference for the selected stars with respect to the mean
cluster value, depicted versus the derived [Fe/H] errors. Stars with a big black open circles
are those in common with \citet{ferraroetal2006}.}
   \label{fig2}
\end{figure*}

The combination of these criteria allow us to remove almost all field stars, as well
as  cluster AGBs. Nevertheless, we counted the number of field
stars distributed in several different regions around the cluster and of an equal cluster
area and found on average that they represent less than 2 percent with respect to the number of selected 
stars located along the delineated RGB ridge line in the $V$ versus $b-y$ CMD. Fig.~\ref{fig2} 
illustrates the distribution of the selected stars in three different CMDs as well as their derived
[Fe/H] values and uncertainties (note that the $V$ vs. $v-y$ panel does not provide independent information, but is included for completeness). 
For comparison purposes we used three arbitrary metallicity-based 
color ranges. As can be seen, selected stars showing no readily visible color spread
in the $V$ versus $b-y$ CMD, do exhibit a clear broadness in the $V$ versus $m_{\rm 1}$ CMD,
which is tightly related to the variation in the cluster metallicity content.
In order to illustrate the impact of the adopted photometric errors in the metallicity
uncertainties, which depend on both the
brightness and the position of a star in the cluster field, we drew with large filled triangles and circles 
stars located closer and farther than 50 arcsec from the cluster's center, respectively.
For completeness, we included all the stars measured within the cluster area and those for
a comparison field region with black and light-blue filled circles, respectively. 
 Four measured AGB stars out of a total of 12 cataloged by \citet{kamathetal2010} are drawn with
a cross. The magenta line is the
theoretical isochrone of \citet{betal12} for an age of 2 Gyr and a metallicity of [Fe/H]= -0.40
dex, shifted by the cluster color excess and distance modulus.
As can be seen, the combination of 
accurate photometry (see Table~\ref{tab:table1} and Fig.~\ref{fig1}) and an
accurate metallicity calibration (see eq.(1)) result in an advantageous tool for
estimating cluster RGB stars' metallicities with uncertainties that, in the case of
the brightest objects, are of the same order than those expected from high-dispersion
spectroscopy.

\section{analysis and discussion}

By using the selected stars, we observe a spread in the mean [Fe/H] values 
derived from the \citet{calamidaetal2007}'s $m_{\rm 1}$ index calibration of $\sim$ 0.35 dex
(see bottom-right panel of Fig.~\ref{fig2}). Nevertheless, in order to
quantify the real spread in the derived [Fe/H] values, we used the well-known
maximum likelihood approach by optimising the probability $\mathcal{L}$ that 
the sample of selected stars with metallicities [Fe/H]$_i$ and errors $\sigma_i$ are 
drawn from a population with mean $<$[Fe/H]$>$ and dispersion W  
\citep[e.g.,][]{pm1993,walker2006}, as follows:\\

$\mathcal{L}\,=\,\prod_{i=1}^N\,\left( \, 2\pi\,(\sigma_i^2 + W^2 \, ) 
\right)^{-\frac{1}{2}}\,\exp \left(-\frac{({\rm [Fe/H]_i} \,- <{\rm [Fe/H]}>)^2}{2(\sigma_i^2 + W^2)}
\right)$.\\

\noindent where the errors on the mean and dispersion were computed from the respective covariance
matrices. If we constrained the sample of selected stars to those with $\sigma$[Fe/H] $<$ 0.15 dex, we 
would obtain W = 0.035$\pm$0.019 dex, while for stars located between $r$ = 50 and 85 arcsec, we would 
derive W= 0.035$\pm$0.023 dex, respectively. By using stars with $\sigma$[Fe/H] $>$ 0.15 dex, any
marginal  dispersion is blurred. We note that an underestimate of the photometric errors would result in a
correspondingly smaller intrinsic spread. For instance, a 20 per cent increase in the adopted
errors reduces the intrinsic spread to W =0.025 $\pm$ 0.020, while a 50 per cent increase leads
to a null W value. This latter result is compatible with using stars with $\sigma$[Fe/H] $>$ 0.15
dex. As for the error in the mean 
[Fe/H], we derived in all the adopted configurations $\sigma$$<$[Fe/H]$>$ = 0.020 dex.  

\begin{figure}
   \includegraphics[width=\columnwidth]{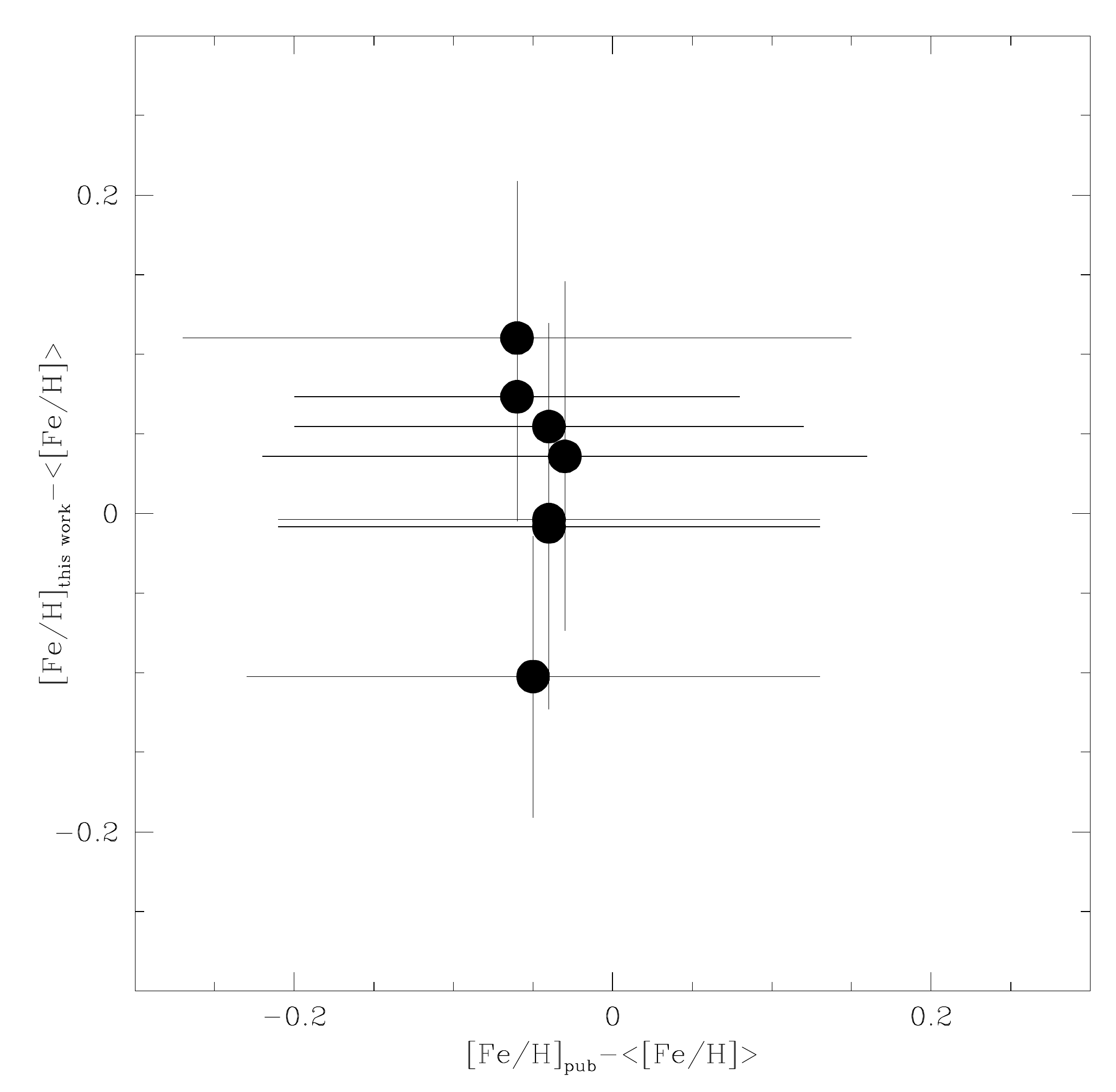}
    \caption{Comparison between \citet{ferraroetal2006}'s metallicities ([Fe/H]$_{\rm pub}$) and 
our values ([Fe/H]$_{\rm this\hspace{0.1cm}work}$).}
   \label{fig3}
\end{figure}

We also computed the mean and dispersion
for 7 stars in common with \citet{ferraroetal2006} (see Fig.~\ref{fig3} and stars with a big black open circle in Fig.~\ref{fig2}) and obtained $\sigma$$<$[Fe/H]$>$ = 0.03 dex and a null W with an upper limit of W $\le$ 0.05 dex, while
from \citet{ferraroetal2006}'s Fe-abundances, we derived $<$[Fe/H]$>$ = -0.43$\pm$0.06 dex and 
a null W with an upper limit of W $\le$ 0.07 dex, respectively.
Although the large uncertainties in \citet{ferraroetal2006} preclude us from seeing a
direct correlation between the present metallicities and those from the high-dispersion
spectroscopy on a star-by-star basis, they are completely consistent.
Since the 7 stars in common do  not span the whole observed [Fe/H] range,
a robust measurement of the small spread is not possible from this small subsample of
stars using either abundance measurement; however, the upper limit derived from this subsample
is consistent with the spread we measure from the full sample.

From the above results, we concluded that
NGC\,1978 exhibit a small spread in the [Fe/H] values, as it is the case of all the
MW old globular clusters, with the exception of 8 with spreads in [Fe/H] $>$ 0.05 dex \citep{carrettaetal2009,ws2012,marinoetal2015}. Nevertheless, since the CN absorption band at
4142\,\AA\, is near to the effective wavelength of the $v$ filter, and hence could be 
measured by the $m_{\rm 1}$ index, the resulting intrinsic [Fe/H] spread could reflect CN variations. 
In such a case, we should interpret this  small spread as reflecting the existence of
MPs. However, there are a number of reasons to favor an interpretation with both light and heavy 
element abundance variation. Firstly, the $m_{\rm 1}$ index was calibrated 
by \citet{calamidaetal2007} as a photometric proxy for the iron abundance. Indeed, \citet{pk2018}
have recently obtained Str\"omgren-based [Fe/H] values for 10 ancient LMC globular clusters 
over a broad metallicity range (-2.1 $\le$ [Fe/H] (dex) $\le -1.0 $) in excellent agreement with mostly high-dispersion spectroscopic values.
Secondly, \citep{limetal2017} showed that old globular clusters with multiple populations
show both light and heavy element abundance variations \citep[see also][]{marinoetal2015}. 
Thirdly, \citet{p18b} showed that accurate Str\"omgren photometry is able to unveil intermediate-age
clusters harboring  MPs. 

\subsection{Self-Enrichment Model}
\citet{bailin2018} presented a model for clumpy self-enrichment in globular clusters, whereby core
collapse supernova ejecta from high mass stars in one protocluster clump are able to enrich
clumps that have not yet formed their stars, resulting in intrinsic metallicity spreads once the clumps
merge to form the final cluster. The model does a good job of matching the 
observed [Fe/H] spreads of MW globular clusters.
An open question is whether the same model can predict metallicity spreads of globular clusters
formed in different environments. In this context, LMC globular clusters like NGC~1978 with
metallicity spreads provide a very useful test.

\begin{figure*}
   \includegraphics[width=\textwidth]{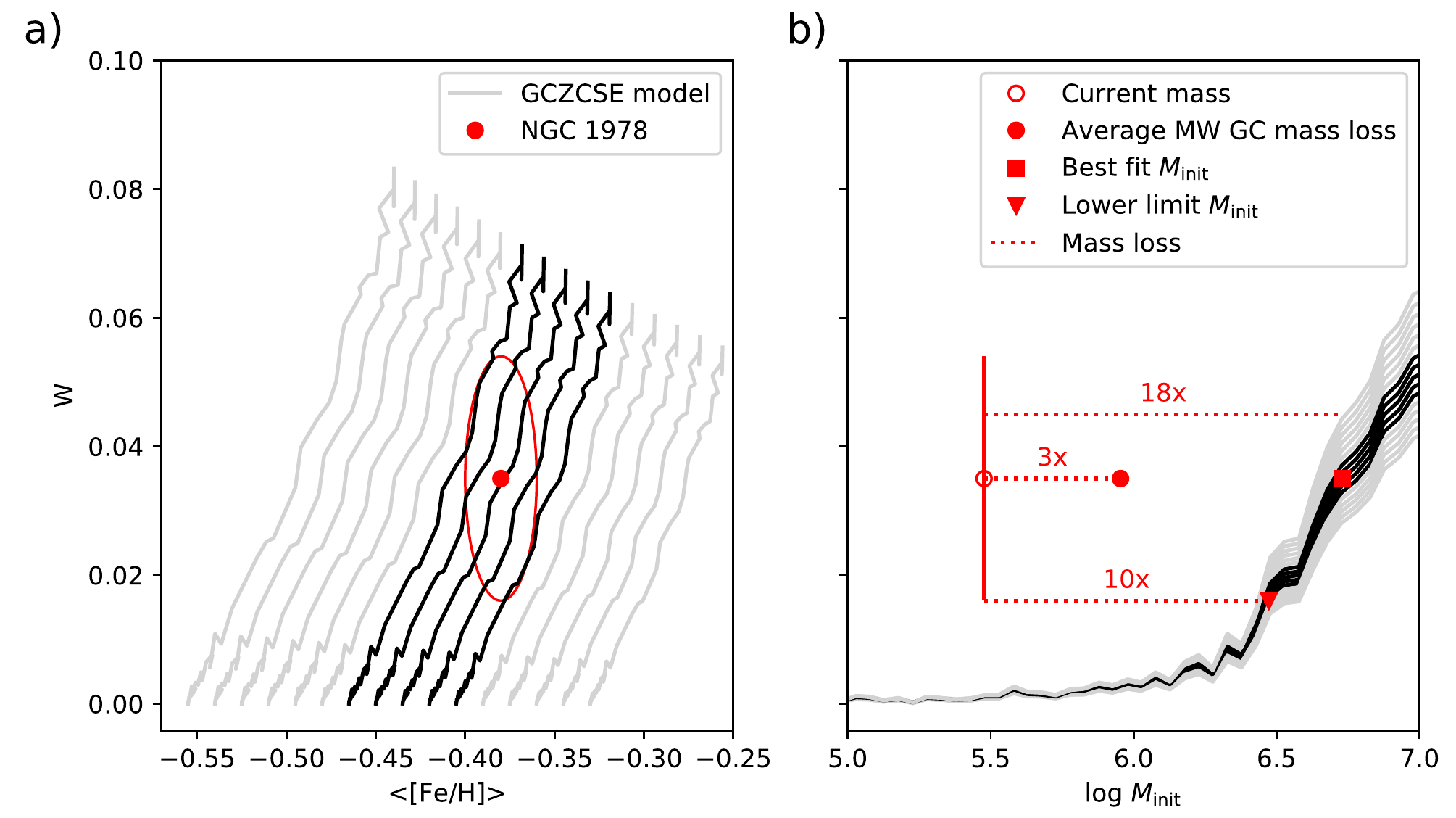}
    \caption{GCZCSE models of clumpy self-enrichment \citep{bailin2018}.
    \textit{(a)} Metallicity spread $W$ as a function of mean
    metallicity $<$[Fe/H]$>$ for a sequence of models with different initial gas cloud
    metallicities. Our observations of NGC~1978 are presented in red, with the ellipse
    indicating the $1\sigma$ uncertainties. Models highlighted in black are those that
    pass through the error ellipse.
    \textit{(b)} The same models are shown as a function of the initial stellar mass
    of the globular cluster. The observed dispersion for NGC~1978 is shown as the open circle
    and error bar,
    plotted at its current stellar mass. The filled circle, square, and triangle indicate
    possible initial values for NGC~1978 if it has had mass loss typical of MW globular
    clusters \citep{bg2018}, if it lies on the best-fit model from panel (a),
    or if it lies on the model track with the minimum value of $M_{\mathrm{init}}$ that
    passes through the error ellipse in panel (a) respectively. The labeled dotted lines
    indicate the required amount of mass loss in each case.}
   \label{fig4}
\end{figure*}

We have run a sequence of models using GCZCSE code \citep{bailin18-gczcse-software}
that result in globular clusters with similar values to those obtained for NGC~1978 (Fig.~\ref{fig4}).
The observed values of
$\mathrm{[Fe/H]}=-0.38 \pm 0.02$ and $W=0.035 \pm 0.019$ are fully consistent with
the \citet{bailin2018} model (panel a), demonstrating that clumpy self enrichment is a viable
mechanism for producing metallicity spreads in LMC clusters.

However,
the metallicity spread in the models is a strong function of metallicity (panel a); although \citet{bailin2018} obtained
models with $W$ as high as 0.40 for typical metal-poor MW clusters,
at $\mathrm{[Fe/H]}\approx -0.38$, the models saturate near $W \sim 0.08$.
This is because spreads in [Fe/H] correspond to the \textit{relative} variation in metallicity:
self-enrichment
causes a fixed spread in the absolute metal mass fraction $Z$, but that spread becomes less and
less noticeable as the magnitude of $Z$ increases. Put another way, a much more massive
protocluster potential well is required to contain sufficient supernova ejecta to achieve
a given value of $W$ at high metallicity. In panel (b), the initial globular cluster
stellar mass of the same models is shown, along with the current observed stellar mass
of NGC~1978 (open circle). The best-fit model from panel (a) requires the cluster
to have had an initial mass $\log M_{\mathrm{init}} = 6.73$ (square), a factor of 18 more than the current mass.
In other words, this model is only viable if mass loss has stripped nearly $95\%$
of its initial mass! The model with the minimum mass that passes through the error ellipse
of the observations has an initial cluster mass of $\log M_{\mathrm{init}}=6.47$ (triangle),
still nearly 10 times larger than its current mass.

Globular clusters do indeed experience significant mass loss during their lives, both
due to stellar evolution effects and tidal stripping. \citet{bg2018} modelled the evolution of
over 150 MW globular clusters, and found a mean mass ratio of 3 between the initial and
final cluster masses (filled circle), with over 10\%\ of clusters having mass ratios
of at least 10. Therefore, clusters with such extreme mass loss are unusual, but not
unexpected. Furthermore, there is evidence that NGC~1978 \textit{is} especially dynamically-evolved.
NGC~1978 lies at a LMC-centric radius of 3.5~kpc;
\citet{pm2018} found that LMC globular clusters within 5~kpc of the center of the LMC
have excesses of extratidal stars indicative of active tidal stripping. Moreover,
they noted six MW clusters whose structural parameters matched those of the inner
LMC clusters; these MW clusters are among the more highly-stripped clusters in
\citet{bg2018}, with a mean mass loss ratio of 6, and extending to over 30 in the case of
Terzan~4.

There are therefore five scenarious that can explain the observed spread in [Fe/H] in NGC~1978:
\begin{enumerate}
 \item The GCZCSE model is a good description of NGC~1978, which has lost at least 90\%\
 	of its mass due to stellar evolution effects and tidal stripping.\label{enum:explanation}
 \item NGC~1978 has experienced less extreme mass loss, but the parameters of the GCZCSE model
 	depend on environment in such a way that LMC clusters experience more self enrichment
 	at a given initial cloud mass.
 	The parameters that could increase $W$ at the relevant masses
 	are: increasing the mixing efficiency, decreasing the duration of star formation,
 	making the stellar initial mass function more top-heavy, and/or decreasing
 	the efficiency of star formation \citep{bailin2018}. However, there is no obvious reason
 	for any of these processes to have operated differently in the LMC than in the MW.
 \item Some of the observed spread in $m_1$ is due to variations in CN abundance rather
 	than [Fe/H]. If the initial cluster mass was only 3 times larger than the
 	present mass, as for the average MW cluster, the predicted spread in [Fe/H] is
 	only $W\approx 0.002$, so in this scenario essentially all of the
 	observed spread must due to this effect.
 \item An unrecognized source of photometric error is adding dispersion to the measurements
 	of different stars in a way that somehow does not show up in our artificial star tests
 	(we consider this unlikely, since we have used the empirical data to estimate the uncertainty).
 \item The success of the GCZCSE model at describing MW globular clusters is a coincidence
 	and is unrelated to the cause of metallicity spreads of globular clusters.
\end{enumerate}

Scenario \ref{enum:explanation}, which is consistent with the model, with observations of
MW globular clusters, and with
the structural parameters of inner LMC globular clusters, is our preferred explanation,
but we cannot presently rule out the others.

\section{conclusion}

With the aim of investigating whether MPs in massive LMC clusters exhibit spreads
in the [Fe/H] values, we made use of accurate Str\"omgren photometry
to estimate individual [Fe/H] values for red giant branch stars in NGC\,1978. Particularly
we used the metallicity-sensitive index $m_1$, along with the high-dispersion spectroscopy-based
metallicity calibration of \citet{calamidaetal2007}, which provided us with [Fe/H]
values with an accuracy between 0.05 dex $\le$ $\sigma$[Fe/H] $\le$ 0.25 dex for previously
selected bonafide cluster members.  

The adopted uncertainties were thoroughly computed by taking into account every source of
error, i.e., those coming from the photometry data set as well as those from the employed 
metallicity calibration. The photometric errors demanded extensive artificial star experiments
in order to be reliably derived. In this sense, we mapped the photometric quality in terms
of both the position of a star from the cluster's center and its brightness. When inspecting
the different CMDs, a clear spread along the cluster's RGB is seen in CMDs that included the $m_1$ 
index, which is related to  a spread in the [Fe/H] values.

In order to quantify such a variation we performed a maximum likelihood analysis, from which
we derived a dispersion of 0.035 dex with an error between 0.019 and 0.023 dex, depending on
whether stars with individual [Fe/H] errors smaller than 0.15 dex, or those
located in the outer cluster's regions are considered. This outcome shows that NGC\.1978 is, 
from the perspective of chemical anomalies in clusters with MPs, similar to the
vast majority of MW globular clusters. On the other hand, we found that the resulting small 
metallicity spread is consistent with a cluster formation model where
self-enrichment of chemical elements is produced by supernova ejecta, provided that a mass loss higher
than 90 per cent is assumed, because of stellar evolutionary effects and tidal stripping.
Nevertheless, other scenarios should not be ruled out, such as a CN abundance variation reflected
on the [Fe/H] values, or less extreme mass loss with a more environmentally dependent self enrichment model.

\acknowledgments

We thank the referee for the thorough reading of the manuscript and
timely suggestions to improve it. 
Based on observations obtained at the Southern Astrophysical Research (SOAR) telescope,
 which is a joint project of the Minist\'{e}rio da Ci\^{e}ncia, Tecnologia, 
Inova\c{c}\~{o}es e Comunica\c{c}\~{o}es (MCTIC) do Brasil, the U.S. National
 Optical Astronomy Observatory (NOAO), the University of North Carolina at Chapel Hill (UNC),
 and Michigan State University (MSU).
 Support for program HST-AR-13908.001-A was provided by NASA through a grant from the
Space Telescope Science Institute, which is operated by the Association for Research in Astronomy, Inc.,
under NASA contract NAS 5-26555.

\bibliographystyle{aasjournal}

\begin{thebibliography}{}
\expandafter\ifx\csname natexlab\endcsname\relax\def\natexlab#1{#1}\fi

\bibitem[{{Bailin}(2018{\natexlab{a}})}]{bailin2018}
{Bailin}, J. 2018{\natexlab{a}}, \apj, 863, 99

\bibitem[{{Bailin}(2018{\natexlab{b}})}]{bailin18-gczcse-software}
---. 2018{\natexlab{b}}, {GCZCSE}, v.1.0.0,  Zenodo, doi:10.5281/zenodo.1253347

\bibitem[{{Balbinot} \& {Gieles}(2018)}]{bg2018}
{Balbinot}, E., \& {Gieles}, M. 2018, \mnras, 474, 2479

\bibitem[{{Bekki}(2017)}]{bekki2017}
{Bekki}, K. 2017, \mnras, 469, 2933

\bibitem[{{Bekki} \& {Tsujimoto}(2016)}]{bt2016}
{Bekki}, K., \& {Tsujimoto}, T. 2016, \apj, 831, 70

\bibitem[{{Bressan} {et~al.}(2012){Bressan}, {Marigo}, {Girardi}, {Salasnich},
  {Dal Cero}, {Rubele}, \& {Nanni}}]{betal12}
{Bressan}, A., {Marigo}, P., {Girardi}, L., {et~al.} 2012, \mnras, 427, 127

\bibitem[{{Calamida} {et~al.}(2007){Calamida}, {Bono}, {Stetson}, {Freyhammer},
  {Cassisi}, {Grundahl}, {Pietrinferni}, {Hilker}, {Primas}, {Richtler},
  {Romaniello}, {Buonanno}, {Caputo}, {Castellani}, {Corsi}, {Ferraro},
  {Iannicola}, \& {Pulone}}]{calamidaetal2007}
{Calamida}, A., {Bono}, G., {Stetson}, P.~B., {et~al.} 2007, \apj, 670, 400

\bibitem[{{Carretta} {et~al.}(2009){Carretta}, {Bragaglia}, {Gratton},
  {D'Orazi}, \& {Lucatello}}]{carrettaetal2009}
{Carretta}, E., {Bragaglia}, A., {Gratton}, R., {D'Orazi}, V., \& {Lucatello},
  S. 2009, \aap, 508, 695

\bibitem[{{Crawford} \& {Mandwewala}(1976)}]{cm1976}
{Crawford}, D.~L., \& {Mandwewala}, N. 1976, \pasp, 88, 917

\bibitem[{{Davidge}(2018)}]{davidge2018}
{Davidge}, T.~J. 2018, \apj, 856, 129

\bibitem[{{Ferraro} {et~al.}(2006){Ferraro}, {Mucciarelli}, {Carretta}, \&
  {Origlia}}]{ferraroetal2006}
{Ferraro}, F.~R., {Mucciarelli}, A., {Carretta}, E., \& {Origlia}, L. 2006,
  \apjl, 645, L33

\bibitem[{{Frank} {et~al.}(2015){Frank}, {Koch}, {Feltzing}, {Kacharov},
  {Wilkinson}, \& {Irwin}}]{franketal2015}
{Frank}, M.~J., {Koch}, A., {Feltzing}, S., {et~al.} 2015, \aap, 581, A72

\bibitem[{{Geisler} {et~al.}(2003){Geisler}, {Piatti}, {Bica}, \&
  {Clari{\'a}}}]{getal03}
{Geisler}, D., {Piatti}, A.~E., {Bica}, E., \& {Clari{\'a}}, J.~J. 2003,
  \mnras, 341, 771

\bibitem[{{Gieles} {et~al.}(2018){Gieles}, {Charbonnel}, {Krause},
  {H{\'e}nault-Brunet}, {Agertz}, {Lamers}, {Bastian}, {Gualandris}, {Zocchi},
  \& {Petts}}]{gielesetal2018}
{Gieles}, M., {Charbonnel}, C., {Krause}, M.~G.~H., {et~al.} 2018, \mnras, 478,
  2461

\bibitem[{{Goudfrooij} {et~al.}(2014){Goudfrooij}, {Girardi},
  {Kozhurina-Platais}, {Kalirai}, {Platais}, {Puzia}, {Correnti}, {Bressan},
  {Chandar}, {Kerber}, {Marigo}, \& {Rubele}}]{goudfrooijetal14}
{Goudfrooij}, P., {Girardi}, L., {Kozhurina-Platais}, V., {et~al.} 2014, \apj,
  797, 35

\bibitem[{{Hauck} \& {Mermilliod}(1998)}]{hm1998}
{Hauck}, B., \& {Mermilliod}, M. 1998, \aaps, 129, 431

\bibitem[{{Kamath} {et~al.}(2010){Kamath}, {Wood}, {Soszy{\'n}ski}, \&
  {Lebzelter}}]{kamathetal2010}
{Kamath}, D., {Wood}, P.~R., {Soszy{\'n}ski}, I., \& {Lebzelter}, T. 2010,
  \mnras, 408, 522

\bibitem[{{Kim} \& {Lee}(2018)}]{kimeal2018}
{Kim}, J.~J., \& {Lee}, Y.-W. 2018, ArXiv e-prints, arXiv:1807.01317

\bibitem[{{Lederer} {et~al.}(2009){Lederer}, {Lebzelter}, {Cristallo},
  {Straniero}, {Hinkle}, \& {Aringer}}]{ledereretal2009}
{Lederer}, M.~T., {Lebzelter}, T., {Cristallo}, S., {et~al.} 2009, \aap, 502,
  913

\bibitem[{{Lim} {et~al.}(2017){Lim}, {Hong}, \& {Lee}}]{limetal2017}
{Lim}, D., {Hong}, S., \& {Lee}, Y.-W. 2017, \apj, 844, 14

\bibitem[{{Marino} {et~al.}(2015){Marino}, {Milone}, {Karakas}, {Casagrande},
  {Yong}, {Shingles}, {Da Costa}, {Norris}, {Stetson}, {Lind}, {Asplund},
  {Collet}, {Jerjen}, {Sbordone}, {Aparicio}, \& {Cassisi}}]{marinoetal2015}
{Marino}, A.~F., {Milone}, A.~P., {Karakas}, A.~I., {et~al.} 2015, \mnras, 450,
  815

\bibitem[{{Martocchia} {et~al.}(2018{\natexlab{a}}){Martocchia},
  {Cabrera-Ziri}, {Lardo}, {Dalessandro}, {Bastian}, {Kozhurina-Platais},
  {Usher}, {Niederhofer}, {Cordero}, {Geisler}, {Hollyhead}, {Kacharov},
  {Larsen}, {Li}, {Mackey}, {Hilker}, {Mucciarelli}, {Platais}, \&
  {Salaris}}]{martocchiaetal2018a}
{Martocchia}, S., {Cabrera-Ziri}, I., {Lardo}, C., {et~al.} 2018{\natexlab{a}},
  \mnras, 473, 2688

\bibitem[{{Martocchia} {et~al.}(2018{\natexlab{b}}){Martocchia}, {Niederhofer},
  {Dalessandro}, {Bastian}, {Kacharov}, {Usher}, {Cabrera-Ziri}, {Lardo},
  {Cassisi}, {Geisler}, {Hilker}, {Hollyhead}, {Kozhurina-Platais}, {Larsen},
  {Mackey}, {Mucciarelli}, {Platais}, \& {Salaris}}]{martocchiaetal2018b}
{Martocchia}, S., {Niederhofer}, F., {Dalessandro}, E., {et~al.}
  2018{\natexlab{b}}, \mnras, 477, 4696

\bibitem[{{Milone} {et~al.}(2009){Milone}, {Bedin}, {Piotto}, \&
  {Anderson}}]{mietal09}
{Milone}, A.~P., {Bedin}, L.~R., {Piotto}, G., \& {Anderson}, J. 2009, \aap,
  497, 755

\bibitem[{{Ordo{\~n}ez} \& {Sarajedini}(2015)}]{os15}
{Ordo{\~n}ez}, A.~J., \& {Sarajedini}, A. 2015, \aj, 149, 201

\bibitem[{{Paunzen}(2015)}]{p2005}
{Paunzen}, E. 2015, \aap, 580, A23

\bibitem[{{Piatti}(2018)}]{p18b}
{Piatti}, A.~E. 2018, ArXiv e-prints, arXiv:1809.08123

\bibitem[{{Piatti} \& {Bastian}(2016{\natexlab{a}})}]{pb16b}
{Piatti}, A.~E., \& {Bastian}, N. 2016{\natexlab{a}}, \mnras, 463, 1632

\bibitem[{{Piatti} \& {Bastian}(2016{\natexlab{b}})}]{pb16a}
---. 2016{\natexlab{b}}, \aap, 590, A50

\bibitem[{{Piatti} \& {Cole}(2017)}]{pc2017}
{Piatti}, A.~E., \& {Cole}, A. 2017, \mnras, 470, L77

\bibitem[{{Piatti} \& {Koch}(2018)}]{pk2018}
{Piatti}, A.~E., \& {Koch}, A. 2018, ArXiv e-prints, arXiv:1809.01709

\bibitem[{{Piatti} \& {Mackey}(2018)}]{pm2018}
{Piatti}, A.~E., \& {Mackey}, A.~D. 2018, \mnras, arXiv:1804.09549

\bibitem[{{Pryor} \& {Meylan}(1993)}]{pm1993}
{Pryor}, C., \& {Meylan}, G. 1993, in Astronomical Society of the Pacific
  Conference Series, Vol.~50, Structure and Dynamics of Globular Clusters, ed.
  S.~G. {Djorgovski} \& G.~{Meylan}, 357

\bibitem[{{Stetson} {et~al.}(1990){Stetson}, {Davis}, \& {Crabtree}}]{setal90}
{Stetson}, P.~B., {Davis}, L.~E., \& {Crabtree}, D.~R. 1990, in Astronomical
  Society of the Pacific Conference Series, Vol.~8, CCDs in astronomy, ed.
  G.~H. {Jacoby}, 289--304

\bibitem[{{Walker} {et~al.}(2006){Walker}, {Mateo}, {Olszewski}, {Bernstein},
  {Wang}, \& {Woodroofe}}]{walker2006}
{Walker}, M.~G., {Mateo}, M., {Olszewski}, E.~W., {et~al.} 2006, \aj, 131, 2114

\bibitem[{{Werchan} \& {Zaritsky}(2011)}]{wz11}
{Werchan}, F., \& {Zaritsky}, D. 2011, \aj, 142, 48

\bibitem[{{Willman} \& {Strader}(2012)}]{ws2012}
{Willman}, B., \& {Strader}, J. 2012, \aj, 144, 76

\end{thebibliography}

\end{document}